\documentstyle[12pt,aasms4]{article}

\begin{document}
\def\gapprox{\mathrel{\vcenter{\offinterlineskip \hbox{$>$}
    \kern 0.3ex \hbox{$\sim$}}}}
\def\lapprox{\mathrel{\vcenter{\offinterlineskip \hbox{$<$}
    \kern 0.3ex \hbox{$\sim$}}}}

\title{Local Magnetohydrodynamical Models of Layered Accretion Disks}

\author{James M. Stone and Timothy Fleming\altaffilmark{1}}
\affil{Department of Astronomy, University of Maryland, College Park, Maryland
20742-2421}
\altaffiltext{1}{current address: Building 600, Fort Monmouth, NJ 07703}

\begin{abstract}

Using numerical MHD simulations, we have studied the evolution of the
magnetorotational instability in stratified accretion disks in which
the ionization fraction (and therefore resistivity) varies
substantially with height.  This model is appropriate to dense, cold
disks around protostars or dwarf nova systems which are ionized by
external irradiation of cosmic rays or high-energy photons.  We find
the growth and saturation of the MRI occurs only in the upper layers of
the disk where the magnetic Reynolds number exceeds a critical value;
in the midplane the disk remains queiscent.  The vertical Poynting flux
into the ``dead", central zone is small, however velocity fluctuations
in the dead zone driven by the turbulence in the active layers generate
a significant Reynolds stress in the midplane.  When normalized by the
thermal pressure, the Reynolds stress in the midplane never drops below
about 10\% of the value of the Maxwell stress in the active layers,
even though the Maxwell stress in the dead zone may be orders of
magnitude smaller than this.  Significant mass mixing occurs between
the dead zone and active layers.  Fluctuations in the magnetic energy
in the active layers can drive vertical oscillations of the disk in
models where the ratio of the column density in the dead zone to that
in the active layers is $<10$.  These results have important
implications for the global evolution of a layered disk, in particular
there may be residual mass inflow in the dead layer.  We discuss the
effects that dust in the disk may have on our results.

\end{abstract}

\section{Introduction}

One of the most promising mechanisms for angular momentum transport in
accretion disks is magnetohydrodynamical (MHD) turbulence driven by the
magnetorotational instability (MRI) (Balbus \& Hawley 1998).  For the
instability to be present, the ionization fraction $x = n_{e}/n_{H}$
(where $n_{e}$ and $n_{H}$ are the number density of electrons and
neutral hydrogen respectively) must be high enough that the neutral-ion
collision frequency is greater than the local epicyclic frequency
(Blaes \& Balbus 1994).  This condition can be met at very small values
of $x$.  For example, at $r=1$~AU in a disk with a density of
$n=10^{13}$~cm$^{-3}$, linear instability requires $x > 10^{-11}$.  The
fact that such low values of $x$ are sufficient to keep the gas and
magnetic field well coupled implies that non-ideal MHD effects are
important only in a few circumstances, such as in protostellar disks
(see, e.g. Stone et al. 2001), or disks in dwarf novae systems (Gammie
\& Menou 1998).

The nonlinear evolution of the MRI in non-ideal MHD has been studied in
several different regimes.  If the ions and electrons are well coupled
to the magnetic field, but the neutrals drift relative to the ionized
species, the plasma is in the ambipolar diffusion regime.  Hawley \&
Stone (1997) used two-fluid simulations to study the saturation of the
MRI in this case.  On the other hand, if collisions are so frequent
that not only are the ions and neutrals well coupled, but also
electrical currents are damped, then the plasma is in the Ohmic
dissipation regime (Jin 1996; Sano \& Miyama 1999).  Saturation of the
MRI in this case has been studied by Sano, Inutsuka, \& Miyama
(1998), and Fleming, Stone, \& Hawley (2000, hereafter FSH), these authors
find sustained MHD turbulence requires the magnetic Reynolds number
$Re_{M} = C_{s}^{2}/\eta\Omega \gtrsim 10^{4}$ (where $C_{s}$ is the
sound speed, $\Omega$ the rotatonal frequency, and $\eta$ the
coefficent of resistivity) for poloidal fields with no net flux.
Finally, if the positive and negative charge carriers can drift
relative to one another, the plasma is in the Hall regime.  The linear
properties of the MRI are modified in unexpected ways in Hall MHD
(Wardle 1999; Balbus \& Turquem 2001); the nonlinear saturation of the
MRI in Hall MHD has recently been studied by Sano \& Stone (2002a;
2002b).

Previous studies of the saturation of the MRI in non-ideal MHD have all
been designed to study the local physics of the instability, and have
therefore used a local approximation termed the shearing box (Hawley,
Gammie, \& Balbus 1995, hereafter HGB) with homogeneous initial
conditions.  In this case, the computational domain represents a small
region deep in the interior of the disk.  However, there are a variety
of important questions that can only be addressed by including the
vertical structure of the disk.  For example, for the case of ideal
MHD, local simulations of vertically stratified disks have been used to
study buoyancy of MRI amplified field (Brandenburg et al. 1995; Stone
et al.  1996, hereafter SHGB) and the formation of strongly magnetized
coronae above weakly magnetized disks (Miller \& Stone 1999).

There are reasons to expect even more interesting vertical structure in
weakly ionized disks.  If the disk is too cold for thermal ionization
of rare earth elements to be important ($T \lesssim 800$~K (Umebayashi
1983)), then the only sources of ionization will be nonthermal, for
example irradiation of the disk by cosmic rays or high-energy photons.
At the interstellar ionization rate due to cosmic rays, $\zeta \approx
10^{-17} ~{\rm s}^{-1}$, the surface layers will be kept sufficiently
ionized that they will be well coupled to the magnetic field (Gammie
1996).  (The presence of small dust grains can significantly change
this picture, however, and lead to much lower ionization fractions,
Wardle \& Ng 1999; Sano et al. 2000.  This will be discussed further in
\S2.2.)  However, the cosmic ray flux is attenuated towards the
disk midplane, meaning the ionization fraction drops significantly,
and non-ideal MHD effects become important.  This has led Gammie
(1996) to propose a ``layered" model for weakly ionized disks, in which
the surface layers are well coupled to the magnetic field and therefore
turbulent (the ``active layers"), while the midplane is decoupled and
queiscent (the ``dead zone").  More recently, Fromang et al. (2002) have explored in more detail the ionization structure of protostellar disks, they
find the extent of the dead zone is extremely sensitive to the
mass accretion rate, the critical magnetic Reynolds number below
which the MRI is suppressed, and the abundance of metals in the disk.

There are a number of interesting implications to Gammie's model.  In
particular, since the accreting layers of the disk have a fixed depth
in surface density, the mass accretion rate increases with the radius
of the disk.  Thus, steady-state solutions are not possible for layered
disks, leading Gammie to suggest that layered accretion could account
for FU Orionis outbursts in T~Tauri disks as accumulated mass in the
dead zone is flushed out by, e.g. gravitational instabilities.  This
idea has been explored further by Armitage, Livio, \& Pringle (2001) by solving
the one-dimensional evolution equation for the disk surface density.
In addition, the layered disk model has important implications for dust
coagulation and planet formation in protoplanetary disks, as the dead
zone may provide a sheltered environment for both processes.

In this paper, we present three-dimensional numerical MHD calculations
designed to explore the local dynamics of layered disks; that is
vertically stratified accretion disks in which the ionization fraction
(and therefore resistivity) varies with vertical height.  We find the
formation of active layers and a dead zone occurs naturally in such a
model, with the critical magnetic Reynolds number found by FSH
providing a good prediction of the demarcation between areas of high
and low angular momentum transport.  We are also able to measure the
rate of mixing between the turbulent layers and the midplane, and how
much heating there is in the midplane due to diffusion of magnetic
energy from the active layers.  Perhaps our most important result is
that we find significant angular momentum transport in the dead layer
by non-axisymmetric density waves driven by turbulent motions in the
active layers.  As discussed in \S5, these results have interesting
implications for the structure and evolution of layered disks.

The paper is organized as follows.  In \S2 we describe our numerical
methods, initial conditions, and the profile of the resistivity adopted
here.  In \S3 we discuss the results of simulations with initially
vertical fields with zero net flux.  In \S4 we discuss the results of
initially azimuthal fields with zero-net flux.  Our conclusions are
given in \S5.

\section{Method}

\subsection{Equations and Initial Conditions}
  
We adopt the shearing box approximation
(HGB) to evolve a local patch of the disk in a Cartesian
frame that is in corotation at a radius $R_{0}$ and angular frequency
$\Omega$.  In this frame, $x, y,$ and $z$ correspond to the radial,
toroidal, and vertical directions, respectively.  In these coordinates
the equations of isothermal MHD may be written as
\begin{equation}
{{\partial \rho} \over {\partial t}} + {\bf \nabla} \cdot (\rho {\bf v}) = 0
\label{continuity}
\end{equation}
\begin{equation}
{{\partial {\bf v}} \over {\partial t }} + {\bf v} \cdot \nabla {\bf v} 
= -{1 \over 
\rho} \nabla \left (P + {B^2 \over 8\pi} \right) + {{{\bf B} \cdot
\nabla {\bf B}} \over 4\pi \rho} - 2 {\bf \Omega} \times {\bf v} 
+ 3 \Omega^2 x {\bf {\hat{x}}} - \Omega^2 z {\bf {\hat{z}}}
\label{momentum}
\end{equation}
\begin{equation}
{\partial {\bf B} \over {\partial t }} = {\bf \nabla} \times [({\bf v} \times 
{\bf B}) - \eta(z) {\bf J}],
\label{induction}
\end{equation}
\begin{equation}
{P=c^2_{s} \rho}
\end{equation}
where $c_{s}$ is the sound speed,  ${\bf J} =(c / 4\pi) {\bf \nabla}
\times {\bf B}$ is the current density, and $\eta(z)$ is the Ohmic
resistivity (which is assumed to be a fixed function of vertical
height), and the other symbols have their usual meaning.  The method
used to compute the vertical profile of the resistivity $\eta(z)$
described in the next section.  The term $-\Omega^2 z {\bf {\hat{z}}} $
in the momentum equation (eq. 2) is the vertical component of the
gravitational field of the central object.  As in SHGB, we use an
isothermal equation of state (equation 4) to prevent an increase in the
scale height of the disk due to heating associated with turbulence in
the nonlinear regime of the MRI.

All of our simulations begin with the disk in a state of hydrostatic
equilibrium.  Thus, the density profile is Gaussian,
\begin{equation}
\rho = \rho_{0}e^{-z^2/H^2}
\end{equation}
where $H^2=2c^{2}_{s}/ \Omega^2$ is the scale height of the disk.  In
order to facilitate comparison with SHGB we adopt the same parameters
and initial conditions as that work.  Thus, we set $\Omega=10^{-3}$,
$H=1$, and $\rho_{0} = 1$, leading to $c_s=7.071 \times10^{-4}$
and $P_{0}=5.0 \times10^{-7}$.

Our computational domain spans $H/2 \leq x \leq H/2$, $0 \leq y \leq
2\pi H$, and $-2H \leq z \leq 2H$.  Our standard numerical resolution
is $60 \times 128 \times 256$ grid points (equivalent to the high
resolution runs in SHGB), except that in order to better resolve the vertical
structure of the disk, we have used twice the number of grid points
in $z$.

We study two types of initial magnetic field configurations: either
initially purely vertical or purely azimuthal fields.  In both cases
the field strength is proportional to $\sin 2\pi x$, so there is zero
net magnetic flux through our computational domain.  The magnetic field
amplitude is parametrized by $\beta$, the ratio of thermal to magnetic
pressure.  At the midplane of the disk, $\beta(0) = P(0) 8 \pi/B^2$,
all of the simulations presented here begin with $\beta(0)= 400$.
The exponential decrease in pressure and density reduces the plasma
parameter $\beta$ as well.  From a local analysis in ideal MHD, the 
critical wavelength of the MRI $\lambda_c =
9.18\beta^{-1/2}$ (HGB), therefore we expect that the most unstable
wavelengths will increase with height.  For azimuthal fields, the field
strength is given a gaussian profile in the vertical direction so that
$\beta$ is initially uniform throughout the volume.

We employ periodic boundary conditions in the azimuthal ($y$)
direction, and either periodic (for initially vertical field runs) or
reflecting (for initially azimuthal field runs) boundary conditions in
the vertical ($z$) direction.  In the radial ($x$) direction,
shearing-periodic boundary conditions are implemented on all runs
(HGB).  Outflow boundary conditions that allow mass, momentum, and
energy to flow out of the computational volume would provide a more
realistic model for accretion disks; however, utilizing outflow
conditions along the $z$ direction is problematic.  As described in
SHGB, when strong ($\beta < 1$) tangled magnetic fields are advected
through free boundaries, field loops are broken.  As the ends of the
loop relax, a large Lorentz force will act on the fluid near the
boundary.  This disturbance is capable of affecting the entire disk.
Because strong, highly tangled magnetic fields are characteristic of
MHD turbulence, we cannot use outflow conditions to study the
nonlinear evolution of the MRI unless the boundary is sufficiently
removed from the disk midplane.  For example, local isothermal
simulations performed by Miller \& Stone (2000) found that at five
scale heights above the disk, a magnetically dominated coronal region
which is stable to the MRI is formed, thus outflow conditions can be
applied there.  Since our simulations do not exceed two scale heights
above the disk, we must adopt either periodic or reflective boundary
conditions.  SHGB found that the choice of vertical boundary conditions did not
significantly alter the vertical structure of the disk.

The initial velocity field is a Keplerian shear flow represented by
$v_{y} = -3\Omega x/2$.  The MRI is seeded with small amplitude,
spatially uncorrelated fluctuations in the velocity field distributed
randomly throughout the simulation volume.

\subsection{Vertical Profile of Resistivity}

Our simulations are motivated primarily by the expected structure of
disks around young stellar objects.  We consider disk radii at which
thermal ionization processes are ineffective.  For protostellar disks,
these regions lie at a radius $\gapprox 0.1 AU$ (Gammie 1996).
We examine the case where the electron fraction is constant with time
and cosmic rays are the only source of ionization, although similar
results are expected if ionization occurs via X-rays instead (Igea \&
Glassgold 1999).

The resistivity is related to the electron fraction $x$ by 
\begin{equation}
\eta=6.5 \times 10^{3} x^{-1} {\rm cm}^2 {\rm s}^{-1}
\end{equation}
(Hayashi 1981), and assuming 
overall charge neutrality, the electron density is determined by 
\begin{equation}
{dn_{e} \over dt} = \zeta n_{H} - \beta x^2 n^{2}_{H} = 0,
\end{equation}
which yields
\begin{equation}
x = \left({\zeta \over {\beta n_{H}}}\right)^{1/2}.
\end{equation}
where $\beta$ is the recombination coefficient.  A standard gas phase
recombination rate due to dissociative recombination is $\beta \simeq
8.7 \times 10^{-6} T^{-1/2}$~cm$^3$~s$^{-1}$ (Glassgold, Lucas, \&
Omont 1986).  At the surface of the disk we assume the ionization rate
due to cosmic rays is the interstellar value, $\zeta_{0} \simeq
10^{-17} s^{-1}$ (Spitzer \& Tomasko 1968).  (We neglect the effect
that magnetized outflows, if present, may have on the flux of cosmic
rays that reach the disk surface.)  However, the flux of cosmic rays is
exponentially attenuated within the disk, with a decay length of
$\Sigma_{CR}= 100{\rm g~cm}^{-2}$.  Thus, the ionization rate within
the disk, $\zeta(z)$, is
\begin{equation}
\zeta(z) \simeq \zeta_{0}e^{-\Sigma(z)/\Sigma_{CR}},
\end{equation}
where the column density $\Sigma(z)=\int^{\infty}_z \rho dz^{\prime}$.
Combining equations (6), (8), and (9) 
and normalizing with respect to disk scale height yields 
\begin{equation}
\eta(z)=\eta_{0}\ {\rm exp}(-z^2/2){\rm exp}\left(
{{\Sigma_0 \over \Sigma_{CR}}{1 \over {2
\sqrt{pi}}} \int^{\infty}_{z}e^{-z^{\prime 2}}dz^{\prime}}\right)
\end{equation}
for an isothermal disk, where $\Sigma_0$ is the total column density
through the disk.

Our expression for the resistive profile (equation 10) has two free
parameters $\eta_0$ (the resistivity at the midplane) and the ratio
$\Sigma_0/\Sigma_{CR}$.  We choose the former so that the magnetic Reynolds
number near the midplane $Re_M \lesssim 10^{3}$, this is below
the critical value found by FSH for sustained turbulence and ensures
the MRI will be damped in the midplane.  The latter we choose to give a
large dynamic range in the resitivity.
Setting $\Sigma_0/\Sigma_{CR} \sim 30$ yields an
$Re_M$ that varies by four orders of magnitude over two scale heights.
Note this implies the disk surface density is 3000 gm~cm$^{-3}$, about
a factor of two larger than the value for the minimum solar nebula.  The range
of values chosen for $Re_M$ was limited by the Courant condition for 
numerical stability.  Choosing values lower than $Re_M=100$ would have placed
a substantial computational burden upon our simulation. Thus it is
worth noting that our midplane values for $Re_M$ are somewhat higher than
those expected in protostellar disks at radii near 1AU.  For example, 
with a number density of $10^{13} cm^{-3}$, T=100 K, and a cosmic ray
ionization rate $\zeta = 10^{-17} s^{-1}$, the fractional ionization is
$x = 10^{-12}$.  This yields a $Re_M \approx 1$.  However, as our
simulations will demonstrate, the values of $Re_M$ considered in this
work are sufficient to stabilize the disk.  Moreover, once the critical
magnetic Reynolds number which stabilizes the disk is reached, we do
not expect the dynamics (such as the Reynolds stress) in the dead zone
to be significantly different at lower $Re_M$.  However, we might
expect that decreasing the value of $Re_M$ at the midplane for our
simulations would result in an increase in the size of the dead zone
and therefore a thinner active layer.

We have neglected effects of dust grains in this analysis.  For a given
ionization rate, the presence of small dust grains can greatly decrease
the ionization fraction since the surface of grains can act as a
catalyst for recombination reactions.  If dust grains with an
interstellar size distribution are mixed throughout the disk, it is
unlikely that even the surface layers of the disk will be sufficiently
ionized to be unstable to the MRI (Sano et al. 2000).  However, dust grains
may grow in size through coagulation, thereby decreasing their surface area
and reducing their effect on the recombination rate.  Alternatively, in
the abscence of turbulence which keeps the grains well mixed
vertically, they may sediment toward the midplane leaving the upper
layers of the disk dust free.  By neglecting the effects of grains, we
restrict the applicability of our results to protostellar disks which
are in the late stages of evolution, where grains have either settled
to the midplane or grown too large to affect the recombination rate.
The calculation of the self-consistent vertical structure of the disk
in the presence of small grains requires following both the non-ideal
MHD of the gas, as well as the dynamics of the grains driven by
aerodynamic forces (to account for gravitational settling and turbulent
mixing); such a calculation is beyond the scope of the current paper.

\section{Vertical Fields with Zero-Net Flux}

Table 1 summarizes the parameters and results of our numerical
simulations.  We first discuss the results from vertical field
simulations, which are labeled by the prefix Z (toroidal field
simulations, which are discussed in the next section, are labelled with
the prefix Y).  At the numerical resolution used here the ratio of the
critical wavelength $\lambda_c$ of the MRI at the midplane (assuming
ideal MHD) to the grid spacing $\Delta z$ ($\Delta y$) for poloidal
(toroidal) fields is 30, indicating it should be captured if present.
(With a finite resistivity, the critical wavelength increases, thus our
estimate is a lower bound on the number of grid points per unstable
wavelength.) Columns (4), (5), and (6) give the time- and
volume-averaged values of magnetic field energy, Maxwell stress, and
Reynolds stress in both the active layers and the dead zone, all
normalized with respect to the midplane pressure.  These averages are
taken from late times, when the active layer has reached saturation.
The location of the boundaries between the active layers and dead zone
is measured from the calculations themselves; as discussed below the
boundaries are generally located where $Re_{M}$ is equal to the
critical value found by FSH.

\subsection{A Model with a small dead zone}

We first discuss run Z1, comparing it to the equivalent ideal MHD run Z4
(run Z4 is identical to run IZ2 in SHGB but with a higher resolution).
The magnetic Reynolds number in Z1 varies from 
$1000$ at the midplane to $5.6 \times 10^{6}$ at $z/H = 2$.
Figure 1 is a ``spacetime" plot of horizontally averaged quantities
$F(z,t)$ versus vertical height and time.  The plot is created by
horizontally averaging each four-dimensional quantity $f(x,y,z,t)$ via
\begin{equation}
F(z,t)={{\int \int f(x,y,z,t)dx dy} \over {\int \int dx dy}} .
\end{equation}
The figure can be compared directly to figure 3 in SHGB.

The linear phase of the MRI occurs over the first three orbits of
evolution.  Suprisingly, the plot of magnetic energy shows the earliest
growth occurs near the midplane, where the resistivity is highest and
the growth rates of the MRI should be lowest.  In fact, this growth is
not associated with the MRI at all, but occurs as a consequence of
reconnection in the initial field topology.  In the highly resistive
midplane, the oppositely directed vertical field components in the
initial conditions reconnect, forming radial fields that are
subsequently sheared by the differential rotation of the disk.  This
shear produces an azimuthal magnetic field which grows linearly with
time.  It is this azimuthal field that produces an increase in the
magnetic energy.  This process does not occur in ideal MHD runs.

The shear amplified field is soon overwhelmed by the exponentially
growing MRI.  The initial saturation of the MRI occurs around orbit 3;
it is evident as a strong peak in the magnetic energy and other
variables at this time.  The highest field energy occurs at $|z/H| <
1.5$ as a consequence of disk stratification.  The decrease in density
with increasing $|z|$ means that the Alfv\'{e}n speed, and also
critical wavelength $\lambda_c$, rise with increasing $|z|$.  The
increase in $\lambda_c$ lowers the peak magnetic field strength at
saturation.  In fact, some of the long wavelength modes near the top
and bottom of the disk may exceed the length of our simulation volume
and thus will not be present.  Near $z=0$ (minimum $Re_M$), the field
energy is reduced, indicating the shorter wavelength modes of the MRI
have been damped by resistivity; only long wavelength modes
develop in the midplane region.  Note the maximum magnetic energy at
orbit 3 has discrete peaks at various vertical positions.  These are
associated with the channel solution that occurs for vertical fields
(Hawley \& Balbus 1991; SHGB).

Beyond the linear phase (after orbit 3) we see the breakdown of the
channel solution via parasitic instabilities (Goodman \& Xu 1994) into
MHD turbulence.  In an ideal MHD run (cf. figure 3 of SHGB), this
turbulence fills the entire volume of the disk.  However, the variation
of the resistivity with height causes a non-uniform structure in the
disk: MHD turbulence is confined to the outer layers alone.  Thus, as
anticipated by Gammie (1996), the disk may be divided into two regions:
two active layers near the surfaces of the disk (characterized by low
resistivity and MHD turbulence), and a dead zone near the midplane
(characterized by high resistivity and a quiescent flow).  The dead
zone forms spontaneasouly after saturation, by orbit 15 it is well
developed.  The boundary between zones is located at approximately $|z|
= z_{B}=0.4$.  At $|z|=z_{B}$ the magnetic Reynolds
number $Re_M \sim 2\times 10^{4}$.  This is in good agreement with the critical
$Re_M$ obtained by FSH for zero mean poloidal fields in uniform disks.
This dead zone persists throughout the remainder of the simulation.
Given the location of the boundaries, and since the vertical profile of
density is little changed from the initial state (eq. 5) (see below),
the column density in the active layers $\Sigma_a$ compared to the dead
zone $\Sigma_d$ follows directly from
\begin{equation}
  \frac{\Sigma_a}{\Sigma_d} = \frac{2 \int_{z_{B}}^{\infty} \exp(-z/H)^{2}}
    {\int_{-z_{B}}^{z_{B}} \exp(-z/H)^{2}} \approx 0.75
\end{equation}

The two layer structure in the disk is evident in the plots of the
Maxwell and Reynolds stress.  Angular momentum transport via magnetic
fields is confined to the active zone.  Interestingly, although the
Reynolds stress is reduced in the dead zone, it is not eliminated
there.  Figure 1 shows that there are irregular increases in Reynolds
stress within the dead zone.  Vertical striations in the plot of
the Reynolds stress are associated with spiral density waves in the disk
(SHGB).
A comparison with the field energy reveals
that the fluctuations in the Reynolds stress correspond with
significant field energy increases in the active layers.  Strong
Lorentz forces in the active zone may be driving these fluctations well
into the midplane region.  We will examine this question in greater
detail below.

The time-averaged vertical structure of the disk is shown in Figure 2,
which plots the density,
plasma $\beta$ parameter, magnetic and kinetic energy, and Maxwell and
Reynolds stress averaged over orbits 15-45.  The density retains
its initial gaussian profile with the same scale height throughout the
nonlinear evolution of the MRI, indicating thermal pressure is still
the dominant force which balances gravity.  Within the dead zone the
effects of magnetic diffusion are clearly evident.  At the midplane MHD
turbulence has been suppressed resulting in an extremely weak
($\beta=1000$) field there.  This is a significant decline in the field
energy as compared to the outer regions; $\beta$ in the midplane is
$\sim$ 100 times greater than in the active layers of the disk.
The maximum value of $\beta$ in the dead zone is $\sim$ 22
times greater than the peak value in the ideal run Z4.  However, in the
dead zone of run Z1, the field rises by an order of
magnitude as the demarcation line, $z=z_B$, is approached.  By $z=1.0$,
$\beta$ has declined to a value comparable to that at the corresponding
point of the ideal run Z4.  This steep increase in field energy is
consistent with the high gradient in resistivity present in the dead
zone.

Tables 1 and 2 list volume averaged values for a number of quantities
in run Z1 in both the active layers and the dead zone.  From the tables
it is evident that the average magnetic energy is three times larger in
the active layer than in the dead zone.  The magnetic energy in the
active layer is confined mostly to the azimuthal component, with
$B^{2}_{y} \sim 7.5 B^{2}_{x} \sim 17.5 B^{2}_z$.  Similar results are
obtained in the ideal run BZ4 and in SGHB.  Azimuthal field also
dominates all other components in the dead zone as well.  In all cases
the background shear favors the growth of the azimuthal component.  The
amplitude of each of the field components in the active layers of run
Z1 is significantly lower (by a factor of at least two) than in the
corresponding locations in the ideal MHD runs.  This is mostly due to
the finite resistivity in the active layers themselves, rather than due
to diffusion into the dead zone.  This is shown by the magnitude of the
Poynting flux
\begin{equation}
F_{P}={\int_s{\eta({\bf J \times B})
\cdot {\bf dA}}}+{\int_s{B^2 \thinspace {\bf v} \cdot {\bf dA}}}
-{\int_s{({\bf v} \cdot {\bf B}) {\bf B} \cdot {\bf dA}}}
\end{equation}
measured at the boundaries between the active layers and
dead zone in run Z1.  The average magnetic energy flux out of the
active regions in one orbit is only $0.3\%$ of the total magnetic
energy in those regions.
We conclude the lower field energy in those regions is a
result of Ohmic dissipation within the active zone itself.

From Table 1, in the active regions the time- and volume-averaged
Maxwell stress exceeds the Reynolds stress by a factor of $\sim 4$;
consistent with the ideal MHD results (run Z4 and SHGB).
However, as with field energy
saturation amplitudes, the stress amplitudes in the active regions are
reduced by about 50\% compared to their values in the same volume of an
ideal MHD simulation due to the finite resistivity in these layers.
Within the dead zone, the Maxwell stress drops rapidly (see Figure 2).
The time- and volume-averaged Maxwell stress in the dead zone is about
1/4 the value in the active layers.  Interestingly, the same effect is
not observed with the Reynolds stress.  The average Reynolds stress in
the dead zone is $\sim 60\%$ of the average Reynolds stress in the
active layers, and $34\%$ of the average Reynolds stress for the ideal
run.  Since the drop in Reynolds stress in the dead zone is not as
significant as the drop in Maxwell stress, the Reynolds stress is now
the dominant mechanism by which transport may occur.  Similar dominance
of Reynolds stress over the Maxwell stress at low $Re_{M}$
was observed in the simulations performed by FSH.

When scaled by the thermal pressure at the midplane, the
total stress in the active layers in run Z1 is $\alpha = W_{xy} / P_0
= 0.006$, whereas in the region within $|z| < 0.4$,  $ \alpha =0.002$
The amplitude of the total stress at the midplane (which is dominated by the
Reynolds stress) is much larger than expected from the uniform resistivity
runs of FSH.  For example, the minimum Reynolds stress (at the midplane)
is about 50 times larger than the Reynolds stress for a uniform resitivity
run with the same $Re_M$ from FSH.
To test whether turbulence in the active layers is driving velocity fluctuations
in the dead zone and producing an enhanced Reynolds stress there, we have
measured the flux of kinetic energy,
\begin{equation}
F_{K}= \int_{S} (\rho v^2) {\bf v} \cdot {\bf dA},
\end{equation}
across the surfaces $|z|= 0.5$.  We find a net flux into the dead zone
which transports $5 \%$ of the average total kinetic energy
into the active layers in one orbit.  This greatly exceeds the magnitude of
the Poynting flux into the dead zone.  

Overshooting of turbulent eddies results in significant mas mixing between the
active and the dead layers of the disk.  The average mixing time (defined as 
the amount of time taken to transport the entire mass of the dead zone across
the active zone boundary) was 3.8 orbits for this simulation.  We find later 
that this time increases as the size of the active layer is decreased.

\subsection{A Model with a Larger Dead Zone}

Run Z2 is identical to Z1 except the resistivity at the midplane is
increased so that $Re_{M}=100$ at $z=0$ and increases to $\sim 5.6
\times 10^{5}$ at $|z|=2$.  This simulation ran for 60 orbits.  The qualitative
evolution of this model is identical to run Z1: a queiscent region is
formed by 15 orbits at the midplane of the disk which is surrounded by
active, turbulent layers.  Once again the boundary between the active
and dead layers is the value of $z$ where $Re_M \sim 2\times 10^{4}$, in this
model this occurs within $|z| < 0.9$.  Thus, the column density of the
active layers compared to the dead zone is $\Sigma_a/\Sigma_d \approx 0.25$,
three times smaller than in run Z1.

The time-averaged vertical structure of the disk in run Z2 is shown in
Figure 3.  Here all quantities have been averaged from orbit 30 to 60,
after saturation has been reached in the active layers.  As with the
other simulations, the gaussian density profile remains constant.  The
lower $Re_M$ in the dead zone has had a significant effect on the field
energy there; the maximum value of $\beta$ has grown by a factor of 8
over the peak for Z1.  The total range in $\beta$ from the dead zone
to the active layers is now over 5 orders of magnitude.  From the plot
of field energy the saturation amplitude in the active layers has been
reduced to $20\%$ of its value in Z1 due to the higher resistivity.
Within the active layers, the azimuthal component of magnetic field
energy dominates by an even greater amount than in Z1; ordering is
$B^2_y \sim 10 B^2_x \sim 27.5 B^2_z$, averaging from orbit 30 to 60.
This is consistent with FSH, where it was found that increasing
resistivity resulted in greater dominance of the azimuthal field.  In
figure 3 we see that the field energy drops off by 2 orders of
magnitude between the active zone and the center of the dead zone.
This is an order of magnitude greater than the drop off observed for
Z1.  The saturation amplitude of the kinetic energy are lower in all
regions of the disk for Z2 as compared to Z1.

An interesting feature of the evolution of run Z2 is the excitation of
vertical oscillations in the disk produced by asymmetrical magnetic
pressures in the active layers.  The two active layers act as
independent dynamos, with local amplification of magnetic field
balanced by resistive dissipation.  There are significant temporal
fluctuations in the total magnetic energy in the active layers
associated with the turbulence.  Since the two active layers are
decoupled, these fluctuations are not correlated, leading to small
differences in the magnetic pressure gradients at the boundary of the
dead zone.  These gradients excite vertical oscillations of the disk
midplane with an amplitude of $0.1H$ and a period of 4 orbits.
The oscillations begin around orbit 10 and continue
for the remainder of the calculation with no significant change in amplitude.
They do not appear to affect the
internal dynamics of the active and dead zones, but might have
interesting implications for the production of winds and outflows from the disk.

The time- and volume-averaged Maxwell stress in the active layers
dominates over the Reynolds stress by $\sim 3.5$.  When scaled by the
thermal pressure, this leads to a total stress of $\alpha =
W_{r\phi}/P_0 =  0.00126$ in the active regions, and $\alpha = 0.00026$
in the dead zone ($|z| < 0.9$).  These represent reductions of $79\%$
and $91\%$ as compared to the active and dead zones of Z1.  The higher
average resistivity in these active layers as compared to Z1 reduces
the saturated stress amplitudes.  For example, the Maxwell stress in
the active regions is $\sim 18 \%$ of the value obtained in the active
layers of Z1, and from $z=1$ to $z=0$ the Maxwell stress drops by
approximately 4 orders of magnitude.  This represents a far greater
drop than occurred in the lower resitive run.  However, much like Z1,
the Reynolds stress does not suffer a similar decline in the midplane.
From Figure 3, we see that the average Reynolds stress within the dead
zone does not vary substantialy from its value in the active layer.  As
with Z1, the velocity fluctuations which give rise to the Reynolds
stress are supported by transport of kinetic energy into this region
from the active layers.  The flux of kinetic energy into the dead zone
during one orbit is $12\%$ of the average kinetic energy within the
active layers.  This transport ratio is larger than that which was
obtained for Z1, due to the depressed saturation amplitudes in Z2.  On
the other hand, the average Poynting flux across the interface between
the active and dead layers ($|z|=0.9$) is insignificant, with the net
average value being only $\sim 0.15 \%$ of the average total magnetic
energy in the active regions.

The very small amplitude of the Reynolds stress might be a result of
purely numerical effects rather than being driven by turbulence in the
active layers.  To test this, we have restarted run Z2 at orbit 45 with
the Lorentz forces removed from the equations of motion (i.e. a pure
hydrodynamics simulation).  This restart ran for 10 orbits.  The effect
is immediate.  Turbulent flow in the active layers, and the Reynolds
stress it produces in the dead zone, quickly decays away on a timescale
of only a few orbits.  This result is in agreement with the Lorentz
force off run performed by HGB.  By orbit 55 (ten orbits after the
restart), the Reynolds stress in the dead zone is 10 times smaller than
when Z2 is continued with MHD.  Thus we conclude that the Reynolds
stress in the dead layer requires MHD turbulence in the active
regions.

Another way to quantify the effect of increasing resistivity on the
dynamics of an accretion disk is to measure the mixing time across an
arbitrary boundary within the disk, defined as the time required for an
amount of mass equal to the total mass within a volume to flow into
that volume.  The mass flux into the volume $|z| < 0.5$ was measured
for runs Z1, Z2, and Z4. (Note that these fluxes are measured at the same
location, although the boundary of the dead zone and active layers is
different in each case.)  Consistent with our measurements of kinetic
energy above, we find that increasing the resistivity decreases the
mass flux into the volume $|z| < 0.5$.  The average mixing time
measured for the ideal run, Z4, was 3.8 orbits.  For the resistive run
Z1, this time increases to 5.8 orbits, and for the highly resitive run
Z2, it is 6.2 orbits.  These results suggest that there is a decreasing
level of interaction between the dead and active layers as resistivity
is increased.  Note that even though we have measured the mass flux at
a location well within the dead zone of run Z2, the mixing time is still
relatively short.

Finally, we have also run a model identical to runs Z1 and Z2, but with
a very large resistivity at the midplane so that $Re_M = 10$ there; this
run is labelled Z3 in Table 1.  In this case the magnetic Reynolds is
so small throughout the entire volume of the disk that MHD turbulence cannot
be sustained.  There is transitory growth of the MRI in the early stages of
evolution, however once the MRI saturates the field dies away and the disk
becomes quiescent.  For this reason the model was only run to 20 orbits; the
time-averaged values of the Maxwell and Reynolds stress during this peroid are
given in Table 1.  In comparison to runs Z1 and Z2, the saturation amplitude
of both quantities at this $Re_M$ is very small.  Based on the results of FSH,
sustained turbulence and transport can only be supported at such low
values of $Re_M$ if the magnetic field has a net poloidal flux; our results
are consistent with this finding.

\section{Azimuthal Fields with Zero-Net Flux}

To examine whether the saturated state of the MRI in a vertically
stratified disk depends on the initial field geometry, we have run two
simulations with a purely toroidal field, labeled Y1 and Y2 in Table
1.  In run Y1, the profile of the resistivity was identical to that
used in run Z1, that is $Re_M=1000$ at the midplane, to $Re_M=5.6
\times 10^6$ at $|z| = 2$.  The resistivity was reduced in run Y2, so
that $Re_M=5700$, at the midplane, to $Re_M = 2.8 \times 10^7$ at $|z|
= 2$.  These simulations were performed at our standard resolution
of $60 \times 128 \times 256$, and used reflective boundary conditions
at the surfaces $z=-2$ and $z=2$.

Figure 4 is a spacetime plot of the horizontally averaged magnetic
energy, Maxwell stress, Reynolds stress, and kinetic energy for run
Y1.   As expected, the linear growth rate of the MRI on purely
azimuthal fields is lower than for vertical fields (Balbus \& Hawley
1992), thus saturation does not occur until orbit 20.  Also note that
shear amplification of radial fields generated by reconnection which is
visible at early time in run Z1 (see Figure 1) does not occur here.

Beyond orbit 15, a dead zone ranging from $z=-1$ to $z=1$ is clearly
evident within the disk.  Despite the same profile of resistivity, this
is approximately twice as large as the dead zone observed in Z1.  The
boundary for the dead zone occurs at $Re_M \sim 3\times 10^{5}$.  This is an
order of magnitude larger than the value of $Re_M$ that marks the
boundary for the dead zone in BZ1.  A similar change in critical $Re_M$
with initial field geometery was observed in FSH;  an order of
magnitude increase in the critical magnetic Reynolds number for
instability for a uniform azimuthal compared to a uniform poloidal
field geometry.  The column density in the active layers compared
to the dead zone is $\Sigma_a / \Sigma_d \approx 0.187$, five times
smaller than in run Z1.

The volume-averaged properties of Y1 are more similar to run Z2 (which
has roughly the same ration of column density in the dead zone to active layers) than run
Z1, which has the same resistivity profile.  This is because azimuthal
fields are more strongly affected by magnetic diffusion than poloidal
fields.  For example, the magnetic energy in the active zone is
approximately 10 times greater than in the dead zone.  The ordering of
field energy components in the active zone of Y1 is $B^2_y \sim 10
B^2_x \sim 33 B^2_z$, which is also similar to that in the
corresponding region of Z2.

Figure 5 plots the vertical profiles of the density, plasma parameter
$\beta$, magnetic and kinetic energy densities, and Maxwell and
Reynolds stress averaged from orbits 30 to 45 for Y1.  Note this is
deep in the nonlinear regime, well after saturation of the MRI.  We see
that the saturation amplitude of field energy in the active zone is $2
\times 10^{-3}$ (normalized by $P_0$) for Y1, which is $13\%$ of its
value for Z1.  Additionaly, the Maxwell stress in the active layers of
Y1 is also only $13\%$ of its value for Z1, leading to $\alpha \sim
8\times10^{-4}$ in the active layers.  In the center of the dead zone
($|z|<0.5$), $\alpha
\sim 5.7 \times 10^{-5}$, a value two orders of magnitude lower than found for
Z1.  Finally, we find that
the kinetic energy in the active region is also reduced by about an
order of magnitude as compared to Z1, due to the decreased levels of
MHD turbulence in the active layers.

The qualitative behavior of the Maxwell and Reynolds stress in these
azimuthal runs is the same as in the poloidal runs; while the Maxwell
stress suffers a substantial decline in the dead zone, the Reynolds
stress is only slightly lower there than in the active region.  In the
active layers, Maxwell dominates over Reynolds stress by a factor of
$\sim$ 4.  This is the same ratio as in the poloidal field runs.
However, within the dead zone, the steep decline of the Maxwell stress
causes it to fall below the Reynolds stress.  The closer we get to the
midplane, the greater the Reynolds stress dominates.  For example,
within the volume enclosed by $|z| < 0.5$, the average Reynolds stress
is 4.7 times larger than the Maxwell stress.  For the volume bounded by
$|z|<0.2$, the Reynolds stress is two orders of magnitude greater than
that observed in Z1 for $|z| < 0.2$.

Run Y2 has a much lower resistivity in the midplane than Y1 discussed
above, so that $Re_M=5700$ there.  The properties of this run are given
in Table 1.  We find the production of only a tenous dead zone in the
early stages of the evolution, which disappears once the active layers
are fully saturated.  Shortly after the onset of the MRI, we find that
within the region $|z|<0.5$, average Maxwell stress is $ \sim 10\%$ of
its value in the outer regions.  However, as the simulation progresses,
the difference between the Maxwell stress in the inner and outer
regions decreases.  By orbit 40, the average Maxwell stress within $|z|
< 0.5$ has climbed to $\sim 60\%$ of its value in the outer regions.
In addition, we find that at later stages in the simulation the Maxwell
stress never falls below the Reynolds stress, even at the midplane.
Correspondingly we find that angular momentum transport rates are high
in the central regions of the disk during the last 20 orbits of the
simulation.  Clearly, the existence of a dead zone depends critically on
whether or not the magnetic Reynolds number in the disk midplane is small.

\section{Summary}

Using numerical MHD simulations, we have studied the evolution of the
MRI in stratified accretion disks including resitivity that varies with
vertical height.  The vertical profile of resistivity is computed
assuming that the dominant source of ionization is external irradiation
by cosmic rays (or perhaps X-rays), so that the ionization fraction
drops dramatically below a penetration depth of $\Sigma_{CR} =
100$~g~cm$^{-2}$.  This profile is appropriate for cold disks around
young stellar objects, or in dwarf nova systems.

The evolution of two different initial field geometries were examined:
a zero net mean poloidal field and a zero net mean azimuthal field.
Additionaly, several different column densities for the disk were
studied (models with larger column densities have a much higher
resistivity at the midplane).  In each case, we find the development of
MHD turbulence is the less-resistive surface layers of the disk with a
queiscent zone in the highly resistive midplane, consistent with the
model of layered accretion put forward by Gammie (1996).  The boundary
between the active layers and the dead zone occurs at the location
where the magnetic Reynolds number $Re_M$ is equal to the critical
value found by FSH for MHD turbulence in the nonlinear regime of the
MRI.  For initially poloidal fields with zero net flux this critical
value is of order $10^{4}$, for initially toroidal fields it is ten
times larger.  This criterion is therefore a useful means by which the
scale of the dead zone may be determined in stratified disks.  Since
the critical value is smaller for fields with net flux, the size of the
dead zone may be correspondingly smaller for that field geometry.

The active regions are charcterized by MHD turbulence and outward
angular momentum transport.  In these regions, as with ideal
simulations, Maxwell stress is the significant source of angular
momentum transport, dominating Reynolds stress by a factor of $\sim
4$.  However, the saturation levels for field energy and Maxwell stress
are less than those obtained in ideal MHD models since Ohmic
dissipation in the active zone is never completely negligable.

Our simulations reveal some unexpected properties of the layered
accretion disk model.  For example, although the MRI is suppressed in
the dead zone the Reynolds stress fails to completely vanish there even
when $Re_M$ is far below the critical value.  Measurement of the flux
of magnetic and kinetic energy from the active layers into the dead
zone supports the idea that velocity fluctuations in the dead zone are
driven by turbulence in the active layers.  These fluctuations drive
non-axisymetric density waves which transport angular momentum (provide
a non-zero Reynolds stress) in the dead zone.  Thus, there is a minimum
level of transport in the dead zone in the presence of active layers.

Our limited numerical resolution allows us to explore only a small
range in the size of the dead zone compared to the active layer; we
have performed simulations in which the ratio of the column densities
of the active to dead layers $\Sigma_a / \Sigma_d$ ranges from 0.187 to
0.75.  However, over this range the Reynolds stress in the midplane
does not seem to depend on the size of the dead zone but rather the
amplitude of the turbulence in the active layers; we find the Reynolds
stress in the midplane is always about 10\% of the Maxwell stress in
the active layers.  For models with very small active layers finite
Ohmic dissipation lowers the saturation amplitude of the MRI, in this
case the Reynolds stress in the midplane scaled by the gas pressure is
$\lesssim 10^{-4}$, whereas in the case of $\Sigma_a / \Sigma_d \approx
0.75$ this ratio was $\gtrsim 10^{-3}$.  Angular momentum transport in
the midplane via driven spiral density waves could have important
implications for the global evolution of layered accretion disks, for
example the timescale for mass accumulation in the dead zone (Gammie
1996) may be changed.

We find significant mass mixing can occur between the active layers and the
dead zone via overshooting of turbulent eddies.  The average mixing time
(defined as the time to transport the entire mass of the dead zone across
the boundary with the active layers) varied from 3.8 to 6.2 orbits as
the size of the active layers decreased.  This mixing may strongly affect
dust settling times in the dead zone.

We also find that turbulence in the active layers can drive global,
vertical oscillations of the disk when the active layers are large.  As
is known from ideal MHD simulations, magnetic field fluctuations in the
tubulece driven by the MRI can be large.  Since each surface layer acts
independently in a layered disk, magnetic pressure fluctuations in the
two layers can be unbalanced, and provide a small amplitude vertical
forcing which drives low-amplitude oscillations.

The flux of energy from the active layers to the dead zone is important
for determining the temperature structure in the disk.  In particular,
if the dead zone is strongly heated by an energy flux from the active
layers, it may become sufficiently ionized that it will couple to the
magnetic field and become active if the temperature rises above $\sim
1000$~K.  We find the Poynting flux from the active layers to the dead
zone to be negligable in all the models considered here.  For example,
in run Z1 the the Poynting flux was only $0.3 \%$ of the field energy
in the active regions.

Finally, it is important to note that we have ignored the effects of
dust in all the models computed here.  Sano et al. (2000) have shown
that even a tiny fraction of small dust grains can significantly alter
the ionization fraction in the disk, and therefore the size of the dead
zone and active layers (see also Fromang et al. 2002).
However, the effect of dust depends strongly
on both the size and spatial distribution of grains.  In particular,
gravitational settling toward the midplane, or coagulation into larger
grains, can reduce the effect of dust.  Since gravitational settling
depends on the degree to which the disk is turbulent, which in turn
depends on the ionization fraction and therefore dust distribution in
the disk, it is clear that self-consistent MHD models of dusty disks
will require following the motion dust grains in concert with the gas
dynamics.

\acknowledgements
We thank Charles Gammie for insightful comments, especially regarding the
calculation of the vertical profile of resistivity.  This work was supported
by the NSF and by the NASA Origins of Solar Systems program.

\newpage

\begin{table}[h]
\caption{Properties of Runs}
\begin{tabular}{cccccc} \hline \hline \\ [-0.3cm]
Run & 
$Re_M$ & 
$Orbits$ &
$\langle \langle B^2/8\pi P_0 \rangle \rangle$ &
$\langle \langle -B_xB_y/4\pi P_0 \rangle \rangle$ &
$\langle \langle \rho v_x \delta v_y/P_o\rangle \rangle $ \\ \hline \\ [-0.2cm]

$Z1$ & 1000 & 45 & 0.015/0.004 & 0.005/0.001 & 0.001/0.0009 \\
$Z2$ & 100  & 60 & 0.003/0.0003 & 0.001/$9 \times 10^{-5}$ & 0.0003/0.0001 \\
$Z3$ & 10   & 20 & 0.001/0.0002 & 0.0006/$7 \times 10^{-5}$ & 0.0001/$9 \times 10^{-5}$ \\
$Z4$ & $\infty$ & 45 & 0.028 & 0.009 & 0.003 \\
$Y1$ & 1000 & 45 & 0.002/0.0002 & 0.0007/$4 \times 10^{-5}$ & 0.0001/$8 \times 10^{-5}$ \\
$Y2$ & 5700 & 60 & 0.012/0.005 & 0.004/0.002 & 0.001/0.001 \\
\end{tabular}
\end{table}

\newpage

\begin{table}[h]
\begin{center}
\caption{Time- and Volume-Averaged Quantities for Run Z1}
\addvspace {0.5cm}

\begin{tabular}{ccccc} \hline \hline \\ [-0.3cm]          
$f$ & 
$<f>$ & 
$<<\delta f^2>>^{1/2}$ &
$<f>$ & 
$<<\delta f^2>>^{1/2}$ \\
\hline \\ [-0.2cm]
$Zone          $ & $Active~~Zone$ & $Active~~Zone$ & $Dead~~Zone$ & $Dead~~Zone$\\
$B^2_x/8\pi P_0$ & $1.7 \times 10^{-3}$ &  $1.5 \times 10^{-5}$ & $2.8 \times 10^{-4}$ & $3.8 \times 10^{-6}$ \\
$B^2_y/8\pi P_0$ & $1.2 \times 10^{-2}$ &  $1.0 \times 10^{-4}$ & $3.8 \times 10^{-3}$ & $3.6 \times 10^{-5}$ \\
$B^2_z/8\pi P_0$ & $7.3 \times 10^{-4}$ &  $7.3 \times 10^{-6}$ & $1.4 \times 10^{-4}$ & $1.6 \times 10^{-6}$ \\
$-B_{x}B_y/4\pi P_0$ & $5.2 \times 10^{-3}$ & $4.0 \times 10^{-5}$ & $1.4 \times 10^{-3}$ & $1.6 \times 10^{-5}$ \\
$\rho v^2_x/2 P_0$ & $2.4 \times 10^{-3}$ & $1.5 \times 10^{-5}$ & $3.0 \times 10^{-3}$ & $3.0 \times 10^{-5}$ \\
$\rho \delta v^2_y/2 P_0$ & $1.8 \times 10^{-3}$ & $1.5 \times 10^{-5}$ & $8.4 \times 10^{-4}$ & $6.7 \times 10^{-6}$ \\
$\rho v^2_z/2 P_0$ & $1.3 \times 10^{-3}$ & $8.5 \times 10^{-6}$ & $1.3 \times 10^{-3}$ & $8.8 \times 10^{-6}$ \\  
$\rho v_x \delta v_y/P_0$ & $1.4 \times 10^{-3}$ & $1.2 \times 10^{-5}$ & $9.0 \times 10^{-4}$ & $1.4 \times 10^{-5}$ \\
\end{tabular}
\end{center} 
\end{table}

\clearpage

\begin{figure}
\epsscale{0.8}
\plotone{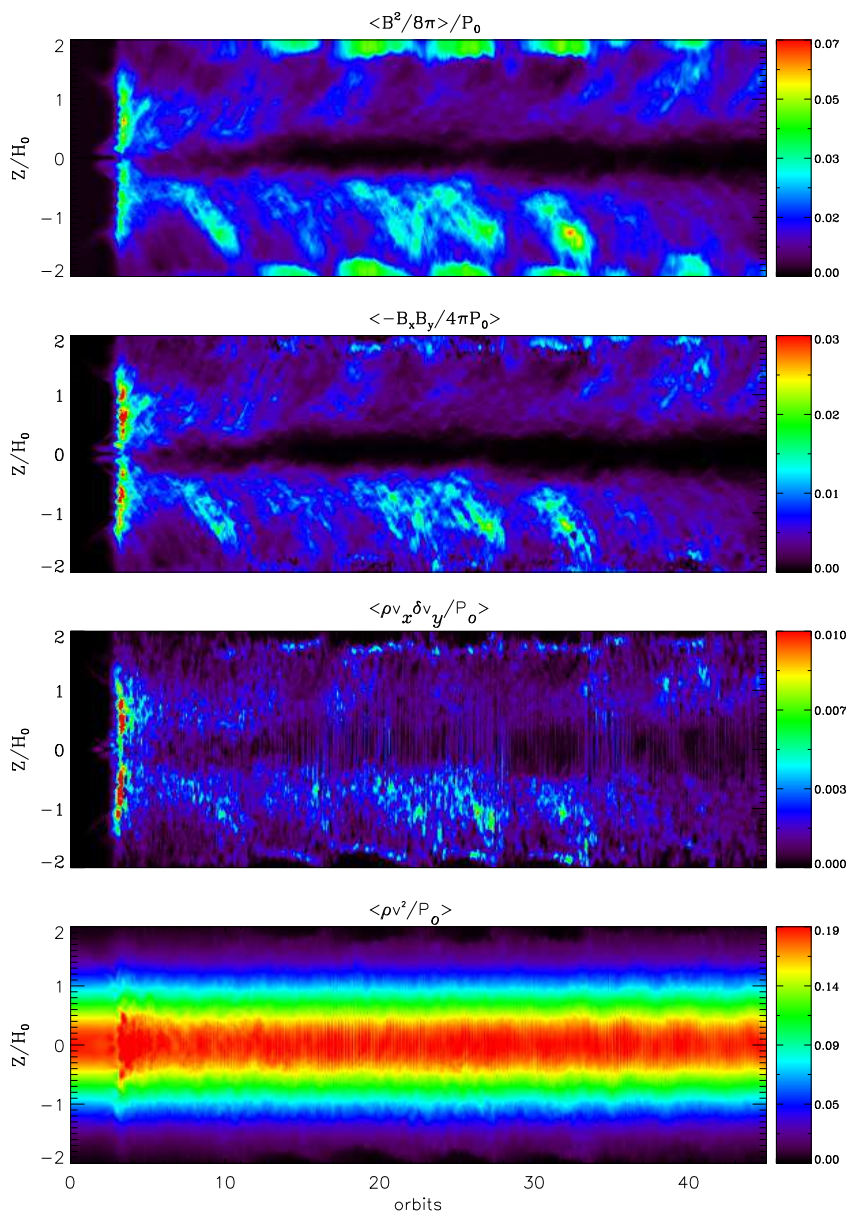}
\figcaption[fig1.ps]
{Spacetime plots of the horizontally averaged magnetic energy, Maxwell
stress, Reynolds stress, and kinetic energy for the zero-net mean field
run Z1. The dead zone is evident by orbit 5: turbulence has ceased
within $|z/H|< 0.4$, however the Reynolds stress is still non-zero in this
region. }
\end{figure}

\begin{figure}
\epsscale{0.8}
\plotone{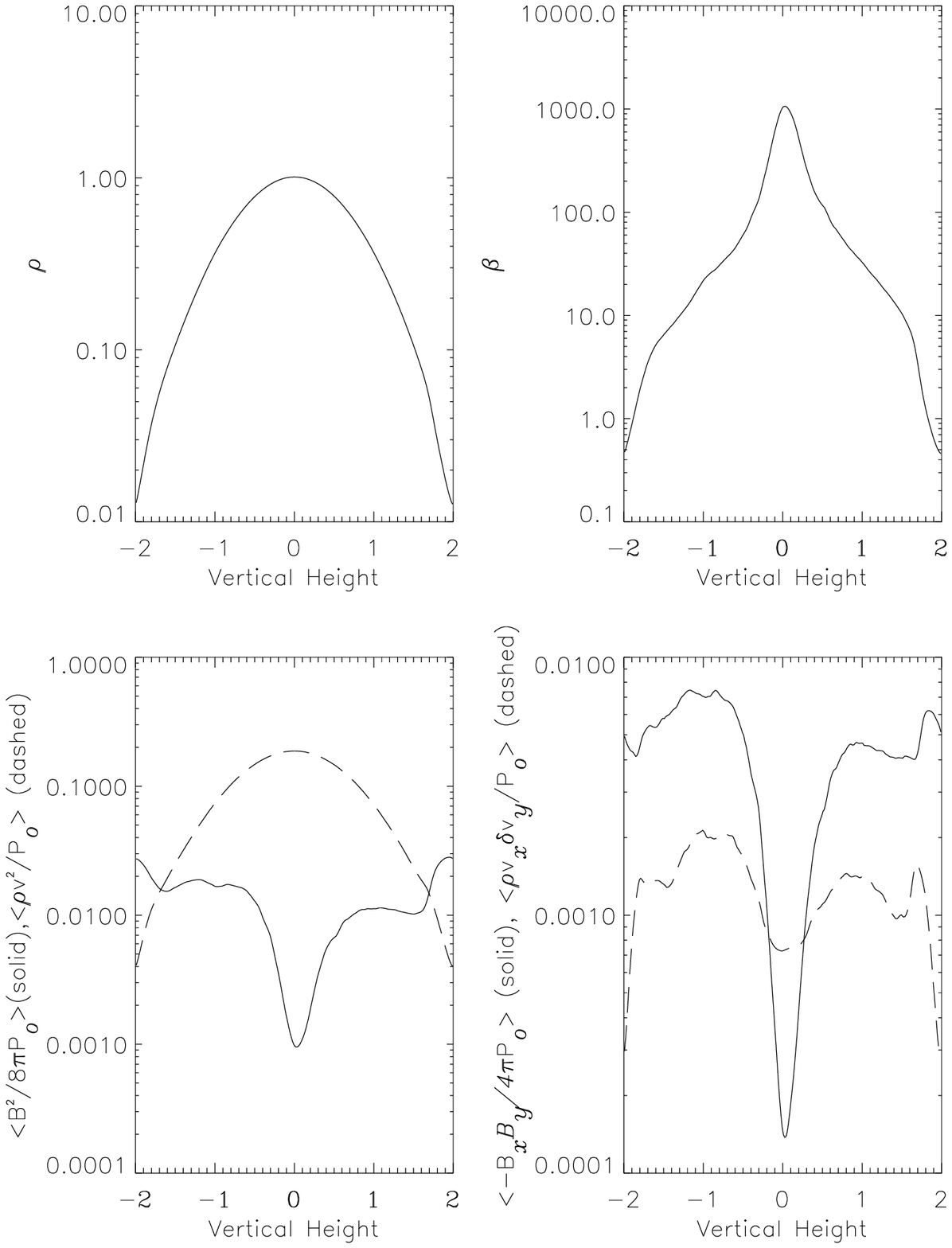}
\figcaption[fig2.ps]
{Time-averaged vertical profiles of the horizontally averaged
density, plasma $\beta$ parameter, magnetic 
(solid line) and kinetic (dashed line) energy densities, and Maxwell 
(solid line) and Reynolds (dashed line) stress in run Z1.  The time-average
is taken over orbits 15 to 45.} 
\end{figure}

\begin{figure}
\epsscale{0.8}
\plotone{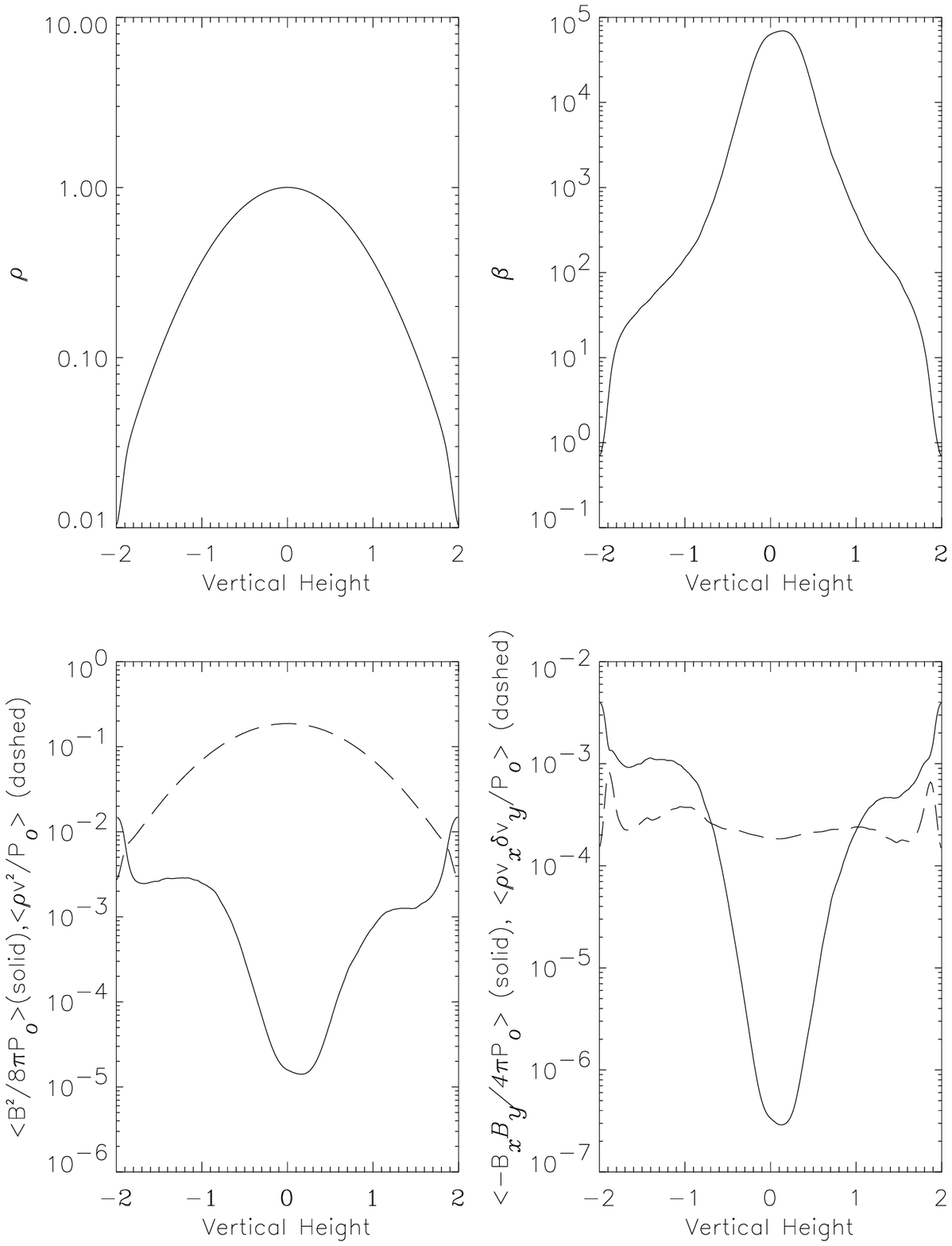}
\figcaption[fig3.ps]
{Time-averaged vertical profiles of the horizontally averaged
density, plasma $\beta$ parameter, magnetic
(solid line) and kinetic (dashed line) energy densities, and Maxwell
(solid line) and Reynolds (dashed line) stress in run Z2.  The time-average
is taken over orbits 30 to 60.}
\end{figure}

\begin{figure}
\epsscale{0.8}
\plotone{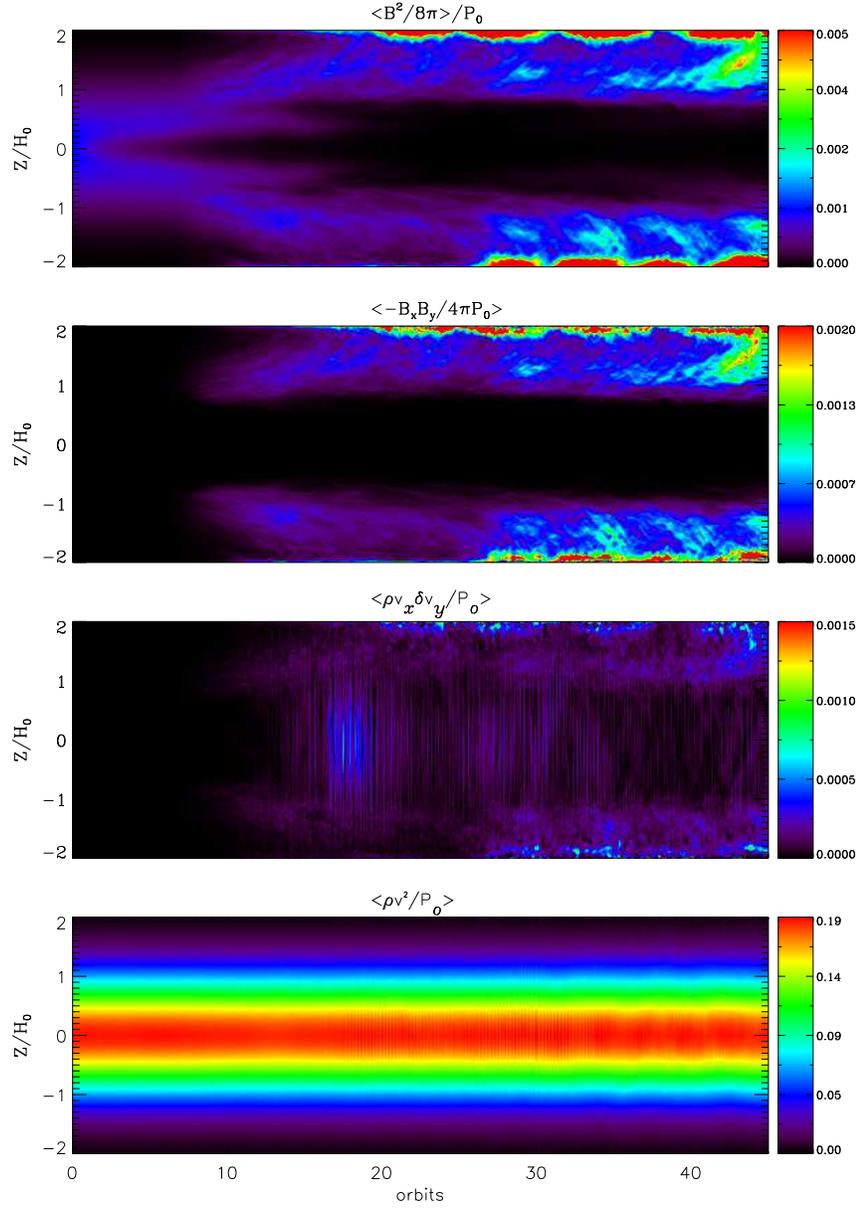}
\figcaption[fig4.ps]
{Spacetime plots of the horizontally averaged magnetic energy, Maxwell
stress, Reynolds stress, and kinetic energy for the zero-net mean field
run Y1}
\end{figure}

\begin{figure}
\epsscale{0.8}
\plotone{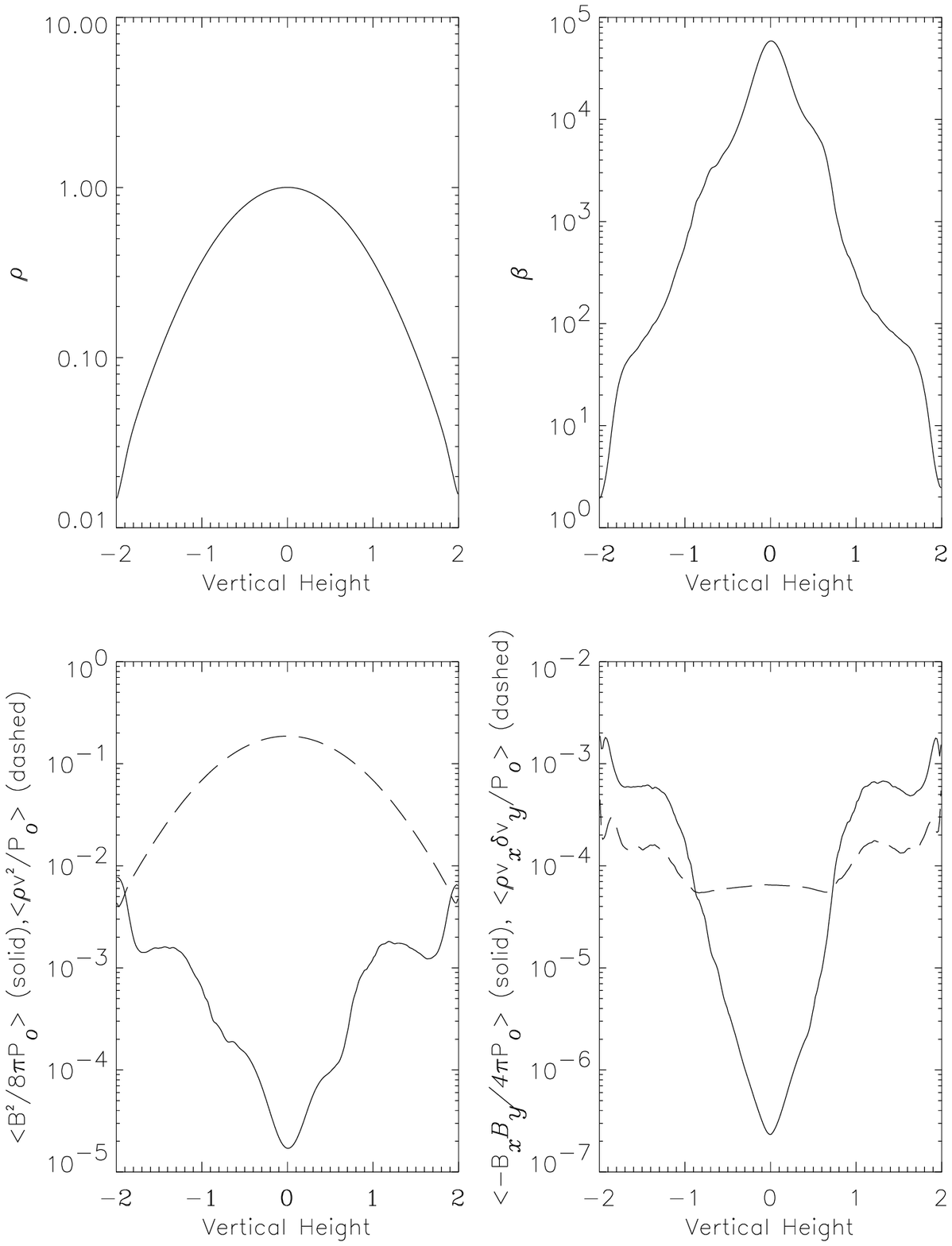}
\figcaption[fig5.ps]
{Time-averaged vertical profiles of the horizontally averaged
density, plasma $\beta$ parameter, magnetic
(solid line) and kinetic (dashed line) energy densities, and Maxwell
(solid line) and Reynolds (dashed line) stress in run Y1.  The time-average
is taken over orbits 30 to 45.}
\end{figure}

\end{document}